\documentclass[apjl,iop]{emulateapj}
\usepackage{amsmath}
\usepackage{amssymb}
\usepackage{graphicx}
\usepackage{parskip}
\usepackage{wasysym}

\shorttitle{WTG Model for Tidally Locked Exoplanets}

\begin{document}
\title{Utility of the Weak Temperature Gradient Approximation for Earth-like Tidally Locked Exoplanets}
\author{Sean M. Mills}
\affil{Department of Astronomy and Astrophysics, The University of Chicago, 5640 S. Ellis Ave, Chicago, IL 60637}
\email{smills@uchicago.edu}
\and
\author{Dorian S. Abbot}
\affil{Department of the Geophysical Sciences, The University of Chicago, 5734 S. Ellis Avenue, Chicago, IL 60637}

\begin{abstract}

  Planets in M dwarf stars' habitable zones are likely to be tidally
  locked with orbital periods of order tens of days. This means that
  the effects of rotation on atmospheric dynamics will be relatively
  weak, which requires small horizontal temperature gradients above
  the boundary layer of terrestrial atmospheres. An analytically
  solvable and dynamically consistent model for planetary climate with
  only three free parameters can be constructed by making the weak
  temperature gradient (WTG) approximation, which assumes temperatures
  are horizontally uniform aloft. The extreme numerical efficiency of
  a WTG model compared to a 3D general circulation model (GCM) makes
  it an optimal tool for Monte Carlo fits to observables over
  parameter space. Additionally, such low-order models are critical
  for developing physical intuition and coupling atmospheric dynamics
  to models of other components of planetary climate. The objective of
  this paper is to determine whether a WTG model provides an adequate
  approximation of the effect of atmospheric dynamics on quantities
  likely to be observed over the next decade. To do this we first tune
  a WTG model to GCM output for an Earth-like tidally locked planet
  with a dry, 1 bar atmosphere, then generate and compare the expected phase
  curves of both models. We find that differences between the two
  models would be extremely difficult to detect from phase curves
  using JWST. This result demonstrates the usefulness of the WTG
  approximation when used in conjunction with GCMs as part of a
  modeling hierarchy to understand the climate of remote planets.

\end{abstract}


\section{Introduction}

In recent years the number of known exoplanets has grown from only a
handful of gas giants to over a thousand planets, including several
roughly Earth-sized rocky planets. There are several ongoing and planned surveys
searching nearby stars for planets including small, Earth-like
habitable zone planets around M dwarf stars, such as MEarth
\citep{2013arXiv1307.3178B}, APACHE \citep{2012MNRAS.424.3101G}, CARMENES \citep{2012SPIE.8446E..0RQ}, and
NGTS \citep{2013EPJWC..4713002W}. This is important because M dwarfs
represent 75\% of the stars in the galaxy and their planets are likely
to be the most easily detectable Earth analogs. Planets near the
habitable zone, with temperatures moderate enough to maintain liquid
water, are of particular interest because water is essential for life
on Earth.

It is likely that planets in the habitable zone of M dwarf stars will
be in spin-orbit resonance, usually in the ``tidally locked''
configuration with one side always facing the star
\citep{1993Icar..101..108K}. Climate modeling studies have shown that
a tidally locked planet with an atmosphere only 10\% of the mass of
Earth's atmosphere is capable of transporting enough heat to the cold
nightside to prevent atmospheric collapse by condensation there
\citep{1997Icar..129..450J, 2003AsBio...3..415J, 2007AsBio...7...30T,
  2007AsBio...7...85S}. Understanding the characteristics of tidally
locked, rocky planet atmospheres is therefore of great
interest. Observational techniques have been proposed to probe the
climate and structure of such planets both by photometry
\citep[e.g.,][]{2009ApJ...703.1884S, 2012ApJ...757...80C,
  cowan12-glint, 2012ApJ...755..101F} and low resolution spectroscopy
\cite[e.g.,][]{1998SPIE.3350..719B, 2000ESASP.451...11F,
  2011AA...532A...1S}.

Tidally locked planets in the habitable zone of M dwarfs should tend
to have a relatively long rotation period (typically tens of days) and
therefore a weak Coriolis force. An analogous situation arises in the
tropics of rapidly rotating planets, such as Earth, because the
horizontal component of the Coriolis force vector is small there. When
the Coriolis force is weak (Rossby Number, $Ro$, $\gtrsim 1$),
advection balances the pressure gradient. For typical wind speeds this
allows only a very weak pressure gradient force and therefore
temperature gradient \citep{1963JAtS...20..607C} and we may reasonably
approximate horizontal atmospheric temperature gradients as zero
(uniform temperature) above the boundary layer, the lower region of
the atmosphere where frictional forces are important
\citep{1995JAtS...52.1784P}.  This situation is referred to as the
``weak temperature gradient'' (WTG) approximation
\citep{sobel2001weak}, and it is brought about by gravity waves.  The
WTG approximation is therefore only valid if the timescale for gravity
wave propagation around a planet is short compared to the radiative
timescale, which excludes its application to extremely hot and/or thin
atmospheres \citep{2013arXiv1306.4673P,2013arXiv1306.2418S}.  3D
general circulation model (GCM) simulations suggest that the WTG
approximation is a useful guiding principle for tidally locked
habitable zone planets \citep{2010JAMES...2...13M}, which leads to
greatly simplified models of the atmospheric dynamics of such planets
\citep{2011ApJ...726L...8P} that can be coupled to models of other
critical processes for planetary climate \citep{2011ApJ...743...41K}.

In addition to elucidating relevant physics, a WTG model can be run
millions of times extremely quickly and could therefore be used in
conjunction with observations to constrain atmospheric
parameters. There are two main advantages to fitting a WTG model to
data as opposed to a simple phenomenological model
\citep{2008ApJ...678L.129C,2013arXiv1304.6398C} or energy balance
model \citep{2011ApJ...729...54C,2013ApJ...766...95L}. First, a WTG
model is based on a dynamically consistent framework so that behavior
constrained by the phase curve is guaranteed to satisfy the relevant
dynamical equations. In contrast, an energy balance model might, for
example, diffuse heat in an unphysical way.  Second, as we will
explain in section \ref{sec:models}, a WTG model includes a solid
surface and solves for the surface temperature, which would be useful
for interpreting phase curves of Earth-like planets.

The goal of this paper is to determine whether a WTG model could be
feasibly distinguished from a GCM using thermal phase curve
photometry, the type of observations relevant to atmospheric dynamics
likely to be available over the next decade. To do this we run a GCM
at a variety of rotation rates, keeping other parameters constant,
then tune a WTG model to the GCM output and compare the thermal phase
curves that each would produce when measured by a remote observer. In
practice a WTG model would be tuned to observations and a GCM requires
many unknown input parameters that would also need to be tuned to
observations, so tuning the WTG model to the GCM is a reasonable
methodology.  We find that in the regime of dry Earth-like $\sim$1 bar
atmospheres, the phase curve a WTG model produces would be nearly
indistinguishable from that a GCM produces for a tidally locked
habitable zone planet orbiting a nearby M dwarf star.  This suggests
that for many situations the WTG approximation is a sufficient
description of atmospheric dynamics for such planets. GCMs would
remain useful for calculating the vertical structure of the
atmosphere, which is critical for radiative transfer, and for cloud
formation and other moisture driven processes, which can strongly
affect the thermal phase curve if present \citep{2013Yang}.

\section{Models}
\label{sec:models}

\subsection{GCM}

The GCM we use is a slightly modified version of the tidally locked
Flexible Modeling System (FMS) from the Geophysical Fluid Dynamics
Laboratory (GFDL) \citep{2011JCli...24.4757M}, which is based on the
model developed by \citet{frierson2006gray} and extended by
\citet{2008JCli...21.3815O}. It solves the 3D primitive equations
(mass continuity, an energy equation, and the Navier-Stokes equations
assuming hydrostatic equilibrium) coupled to a grey gas longwave
radiative transfer scheme. The main difference between the model we
use and that of \citet{2011JCli...24.4757M} is that we set surface
evaporation to zero so that we consider a dry atmosphere. We set the
incident stellar flux to 800 W/m$^2$ ($\approx$0.6 S$_\odot$) in the
simulations we present, but find similar results for higher values,
and assume a constant surface albedo of 0.38, which yields an
effective top-of-atmosphere (TOA) albedo of $\approx$0.30 due to
atmospheric absorption of stellar radiation. We set the linearized
grey gas optical depth to 1.2 and the atmospheric pressure to 1.0
bar. We vary the Coriolis force to simulate different rotation periods
ranging from 1 day to 1 year. For comparison, a tidally locked planet
around an M1 star with a $\sim$50 Earth day tidally locked year would
receive the same stellar flux as the Earth does from the sun, and a
planet with a $\sim$2 day orbit would receive the same flux around an
M6 star. Each simulation is 1000 days long and we average variables
over the final 100 days. The simulations generally reach statistical
steady state in roughly 300 days.

\subsection{WTG Model}

Our treatment of the WTG approximation for a tidally locked planet is
similar to that of \citet{2011ApJ...726L...8P}. We restrict our
investigation to a dry, rocky planet with zero eccentricity. We do not
consider moist effects such as shortwave absorption by atmospheric
water vapor, clouds, and latent heat release.

The WTG model is one dimensional horizontally and has two layers in
the vertical. The model uses surface and top of atmosphere energy
balance to determine the atmospheric temperature ($T_a$) and surface
temperature ($T_s$).  By construction the atmospheric temperature is a
constant, but the surface temperature depends on the angle from the
terminator ($\theta$, which ranges from 90 degrees at the substellar
point, to 0 at the terminator, to -90 at the antistellar point). This
model cannot capture axially asymmetric effects, such as advection of
an atmospheric hot spot by equatorial superrotation. While this type
of effect is important for hot Jupiters
\citep{showman2002atmospheric,knutson2007map}, it is unlikely to be a
zeroth order effect for habitable zone terrestrial planets.

Surface energy balance yields
\begin{equation}
(1 - \alpha) S[\theta] + e_a \sigma T_a^4 =  \sigma T_s[\theta]^4 + a \cdot (T_s[\theta] - T_a),
\label{eq:surf-bal}
\end{equation}
where 
$S[\theta] =\biggl\{ \begin{aligned}
    F_{\star} \sin \theta,\ & \forall \theta>0\\
    0,\ & \forall \theta<0
  \end{aligned}$, $F_{\star}$ is the incident stellar flux at the
  substellar point, $\alpha$ is the albedo, $\sigma$ is the
  Stefan-Boltzmann constant, $e_a$ is the atmospheric longwave
  emissivity, and $a$ is the surface-to-mid-troposphere exchange
  constant, which implicitly includes sensible heat flux by dry
  turbulent exchange and atmospheric convection.  We have assumed that
  the atmosphere is transparent to solar radiation, but has some
  longwave opacity. We take the surface albedo to be a constant and
  the surface emissivity to be one. We set $a$ to zero wherever $T_a >
  T_s$ since, to a first approximation, stable stratification prevents
  sensible heat exchange and convection. Finally, the global mean
  energy balance at the top of atmosphere can be written as
\begin{equation}
\int_{-90}^{90} (1 - \alpha) S[\theta] \cos \theta \mathrm{d}\theta = \int_{-90}^{90} ((1-e_a)\sigma T_s[\theta]^4+ e_a \sigma T_a^4) \cos \theta \mathrm{d}\theta.
\label{eq:toa-bal}
\end{equation}

Equations (\ref{eq:surf-bal}) and (\ref{eq:toa-bal}) can be solved for
the constant atmospheric temperature ($T_a$) and the surface
temperature as a function of theta ($T_s(\theta)$) if $\alpha$,
$\epsilon_a$, and $a$ are specified\footnote{Note that we assume here that
  $F_\ast$ is known. If $F_\ast$ is not known, then $F_\ast(1-\alpha)$
  should be treated as a single unknown parameter so that the model
  still has three tunable parameters.}. We choose the albedo ($\alpha$)
of the WTG model so that it matches the global mean TOA albedo of the
GCM. We then choose $\epsilon_a$ and $a$ to minimize the least square
distance between the WTG model and the GCM for the following
parameters: the TOA upward longwave flux at the substellar point, the
TOA upward longwave flux averaged over the nightside, and the surface
temperature at the substellar point.

\section{Results}

\begin{figure}[h!]
\centering
\includegraphics[scale=.42]{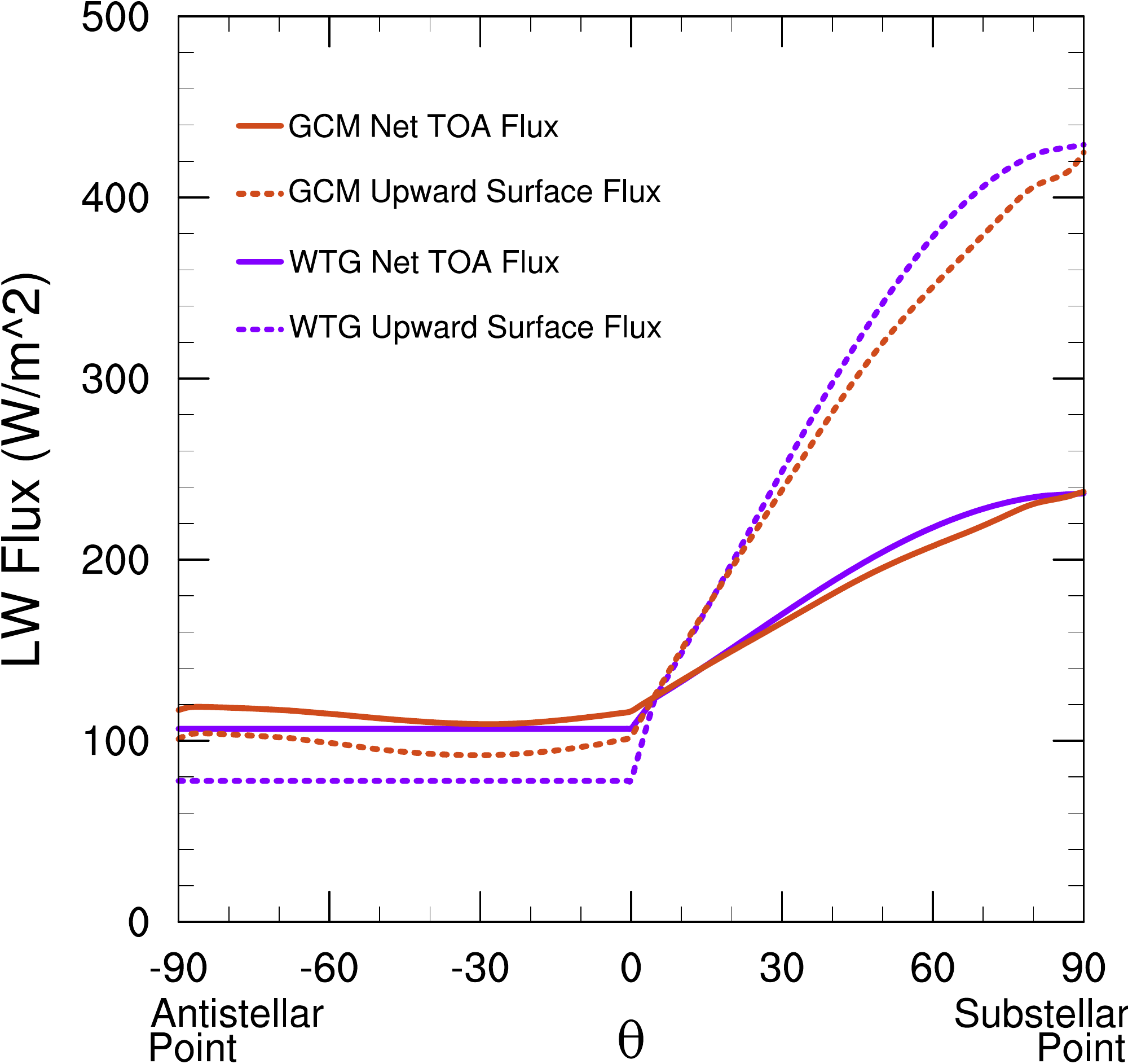}
\caption{This plot demonstrates the similarity of the surface and
  top-of-atmosphere upward longwave flux profiles of the GCM and a WTG
  model tuned to it. $\theta$ is the angle from the terminator. The
  GCM is in a 10 day tidally locked orbit and converged results are
  averaged spatially so that they can be plotted as a function of
  $\theta$.}
\label{fig:comparison}
\end{figure}

\subsection{Comparison of  WTG Model to GCM}

Inspection of the GCM results indicates that the WTG approximation is
reasonable. For example, when the rotation period is set to five days,
the temperature at a given pressure level in the mid-troposphere
($\approx$0.5 bar) varies by less than 10~K around the planet (not
shown). As a result, the WTG model can be easily tuned to reproduce
the broad features of the GCM (Fig.~\ref{fig:comparison}), despite the
fact that it has only three tunable parameters. The similarity between
the longwave fluxes in the WTG model and the GCM when plotted as a
function of the angle from the terminator is striking, and emphasizes
the excellent performance of the WTG approximation. 

It is useful to confirm that the WTG parameters after the fit are
physically reasonable and comparable to those in the GCM.  The albedo
in the WTG model is set to exactly match the average top-of-atmosphere
albedo of the GCM, so there is a direct correspondence between the
models. The emissivity, $e_a$, is a parameter meant to impart the
gross behavior of longwave radiation into the WTG model. The fact that
the net ``greenhouse forcing,'' or difference between surface and
top-of-atmosphere longwave fluxes, is similar between the GCM and WTG
model (Fig.~\ref{fig:comparison}) confirms that the fit to $e_a$ is
reasonable. The final WTG parameter, $a$, the
surface-to-mid-troposphere exchange constant, cannot be easily
compared to a particular parameter in the GCM because it accounts for
the combination surface turbulent exchange and atmospheric convection.

\begin{figure}[h!]
\centering
\includegraphics[scale=.42]{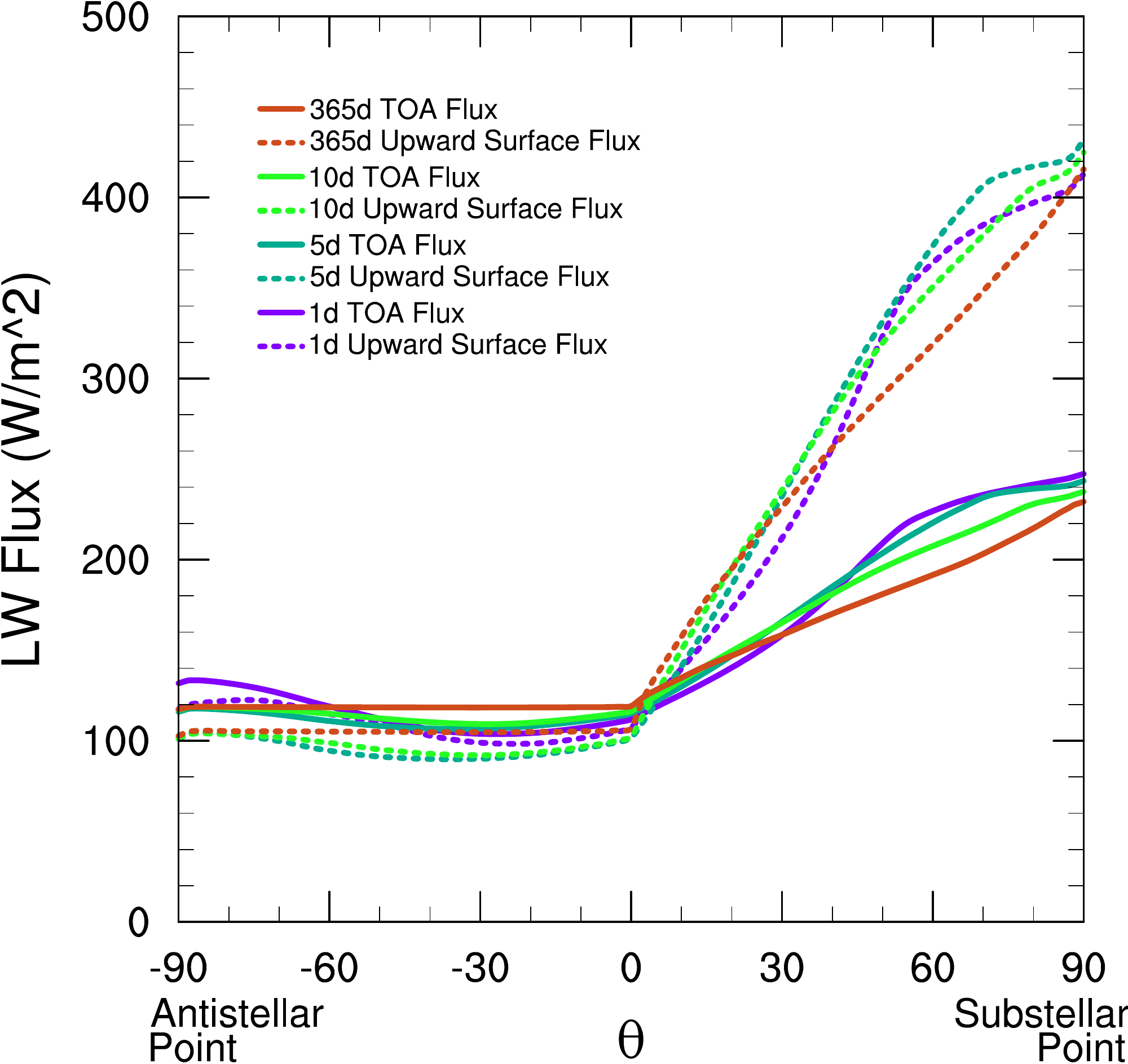}
\caption{A comparison of the upward surface (dotted) and
  top-of-atmosphere (TOA, solid) longwave radiation flux as a function
  of angle from the terminator at different rotation rates in the
  GCM. There is only slight variation in the radiative fluxes when the
  rotation period is varied over orders of magnitude.}
\label{fig:olrcompare}
\end{figure}

Consistent with \citet{2011JCli...24.4757M}, the rotation rate has
remarkably little effect on TOA and surface upward longwave flux
profiles in the GCM when they are plotted as a function of the angle
from the terminator (Fig. \ref{fig:olrcompare}). This is likely due to
the fact that even at high (Earth-like) rotation rates, the tropics,
which dominate emission, still obey the WTG approximation. Others have
observed a dynamical transition when the Rossby radius falls below the
planetary radius at rotation periods less than a few days
\citep{2011Icar..212....1E, 2013AA...554A..69L,2013Yang} that leads to
increased equatorial superrotation (positive zonal wind at the
equator). We observe this as well, but find that even in this regime
the longwave flux profiles do not deviate much from those at low
rotation rates (Fig. \ref{fig:olrcompare}).

\subsection{Difficulty of Observing Model Differences}

Measurement of variations in the disk-integrated broadband thermal
phase curve due to a planet's hot dayside periodically facing towards
and away from Earth is the main tool we have to gain insight into the
atmospheric dynamics of tidally locked exoplanets.  An example of a
major success of this technique is its use on the hot Jupiter
HD 189733b \citep{knutson2007map, 2009ApJ...690..822K} to confirm the
prediction \citep{showman2002atmospheric} of equatorial superrotation
and eastward advection of the hot spot that results from strong
heating on the dayside and cooling on the nightside
\citep{ShowmanandPolvani2011}. Although the Earth-like planets under
consideration would be far less luminous and emit further in the
infrared, their phase curves would still likely be observable
sufficiently close to Earth ($\lesssim$$20$~pc) with next generation
telescopes \citep{2011AA...532A...1S,2012ApJ...757...80C,2013Yang}.

As a test of the WTG approximation, we compare the thermal phase
curves that the GCM and WTG model would produce for a distant observer
using the methodology of \citet{2013Yang}, considering different
observer inclination angles. We calculate the phase curve as the
variation in total thermal flux that would be observed as the planet
orbits the star assuming a stellar radius $R_\ast=0.2 R_\odot$, a
stellar temperature $T_\ast=3000$~K, a planetary radius
$R_p=R_\oplus$, and a planetary emission temperature
$T_{ef}=240$~K. We assume that the Mid-Infrared Instrument (MIRI) on
the James Web Space Telescope (JWST) will be the most advantageous
instrument in the near-term, and assume that we could integrate the
thermal flux from 10-28~$\mu$m using this instrument. 

\begin{figure}[h!]
\centering
\includegraphics[scale=.42]{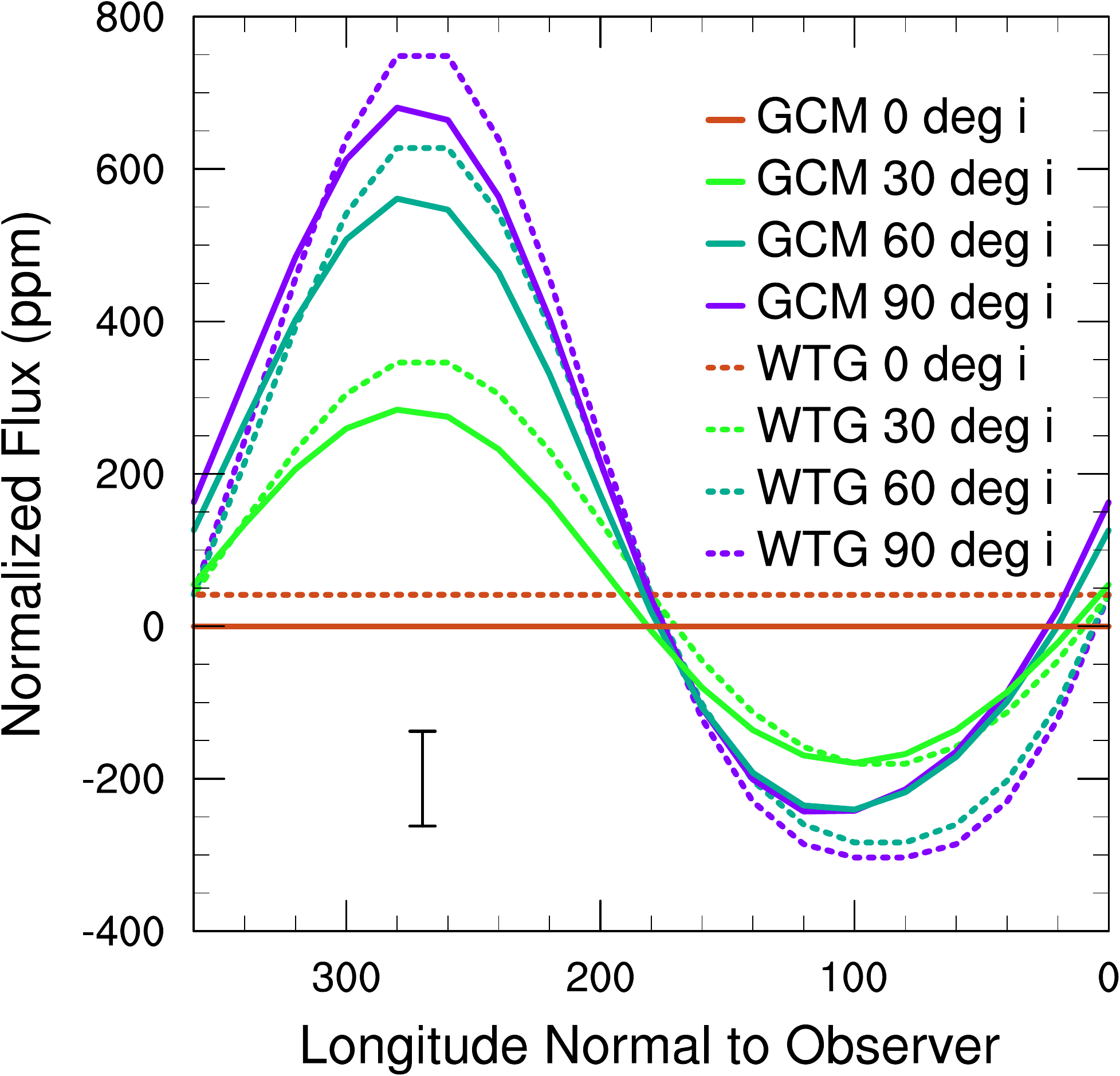}
\caption{A comparison of the predicted phase curves of the GCM run
  with a 10 day orbital period and the WTG model. The horizontal axis
  is the phase ($\frac{360\times [time]}{[Orbital\ Period]}$). The
  vertical axis is the deviation (in ppm) of the total (including both
  star and planet) infrared flux from its mean value. The error bar
  represents the approximate expected precision of a one day
  integration using the MIRI instrument on JWST for an Earth-sized
  planet at a distance of 20~pc. Differences between the WTG model and
  the GCM would not be detectable at this distance.}
\label{fig:phasecurve}
\end{figure}

In order to calculate the expected precision, we scale the photometric
precision of $4\times10^{-5}$ obtained by \citet{2012ApJ...754...22K}
using the Spitzer Space Telescope, which is about 50\% larger than the
photon noise. Phase curve measurements will be easiest to interpret
for a transiting planet, so we assume a distance appropriate for the
nearest transiting tidally locked habitable zone terrestrial planet. A
conservative estimate of 0.04 habitable zone terrestrial planets per M
dwarf yields an expected distance of 20~pc, which we will adopt as our
fiducial value. We also consider distances of 5~pc and 10~pc because
different estimates of planetary frequency
\citep{2013arXiv1302.1647D,2013arXiv1303.3013M} and habitable zone
width \citep{2013ApJ...767L...8K,2013Yang} yield 0.5--1 habitable zone
terrestrial planets per M dwarf. Finally, we assume an integration
time of one day.

Assuming a transiting planet at 20~pc yields a precision of 129~ppm,
at which the phase variations of a dry Earth-sized terrestrial planet
with a 10 day orbital period should be observable at 7-$\sigma$
precision (Fig.~\ref{fig:phasecurve}). The maximum difference between
the WTG model simulation and the GCM simulation would be entirely
undetectable at this distance (Fig.~\ref{fig:phasecurve}). Even if the
planet were at 5~pc (32~ppm precision), the maximum difference between
the WTG model and GCM would still probably be undetectable
(2.5-$\sigma$). Given that all detection significances would increase
linearly with planetary radius, the difference between the WTG model
and the GCM would only start to be detectable for a large super-Earth
at a distance of 5~pc (assuming that this difference does not scale
strongly with planetary radius and that the atmosphere is not
extremely thin).

\begin{figure}[h!]
\centering
\includegraphics[scale=.355]{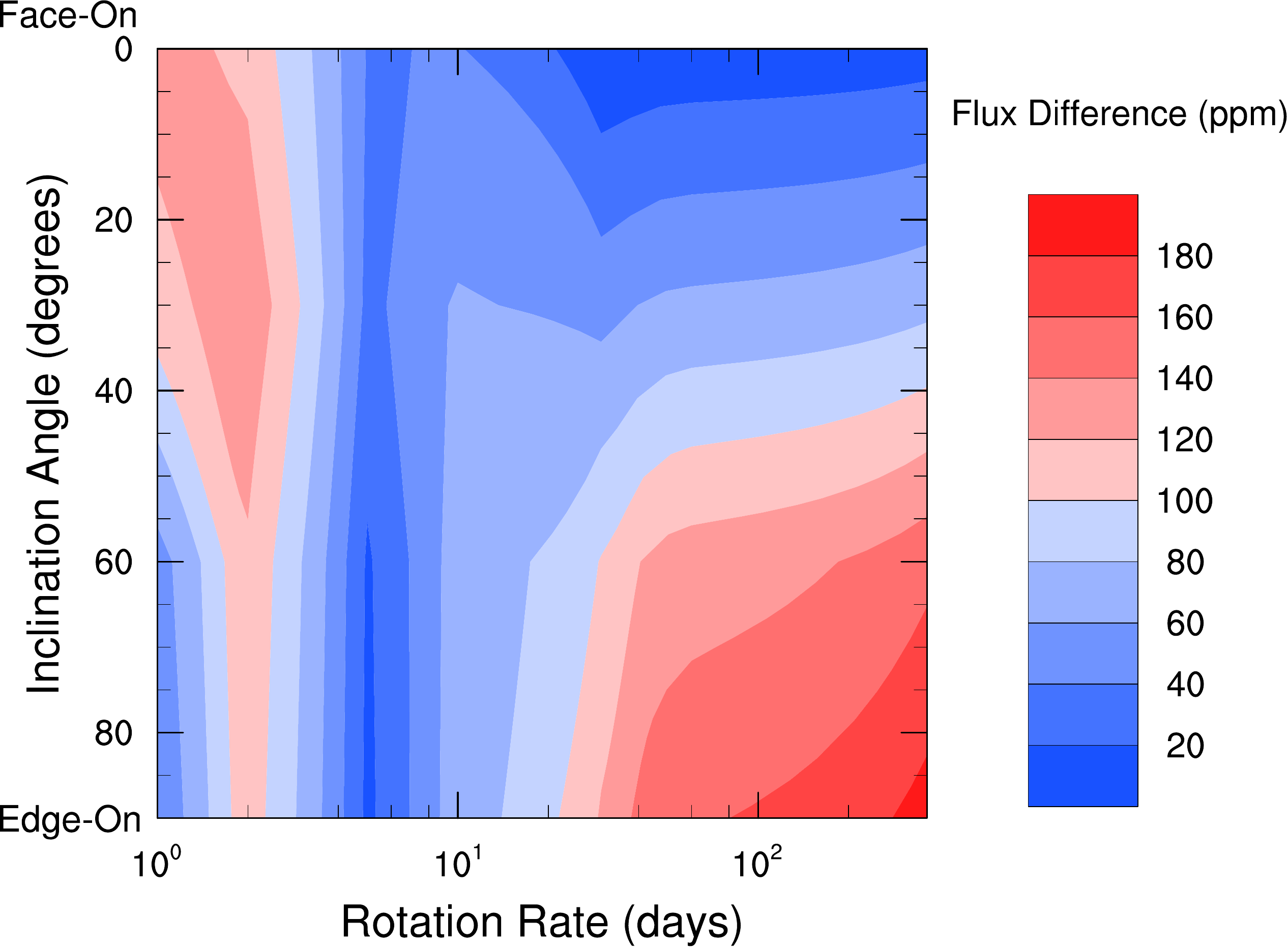}
\caption{The maximum difference between phase curve that the WTG model
  and the GCM would produce assuming an Earth-sized planet orbiting an
  M star with radius 0.2$R_\odot$. The difference is plotted in ppm
  and should be compared to an instrumental precision for one-day
  integrations using MIRI instrument on the JWST of 129~ppm if the
  system is at a distance of 20~pc, 65~ppm at 10~pc, and 32~ppm at
  5~pc.}
\label{fig:contour}
\end{figure}

When we vary the planetary rotation rate and inclination angle, we find
that the maximum difference between the WTG model and the GCM would
generally not be detectable (Fig.~\ref{fig:contour}). The most
important exception is for relatively low rotation rates (long period
orbits) in near transiting configuration. The difference between the
WTG model and the GCM might be detectable for these planets,
particularly if they are at distances of less than 10~pc. It is
interesting that the WTG gradient approximation actually becomes less
effective at lower rotation rates. This is due to the sharper profiles
of infrared emission to space as a function of angle from the
terminator at lower rotation rates (Fig.~\ref{fig:olrcompare}), which
are hard for the WTG model to fit.

\section{Discussion and Conclusions}

We find that a WTG model provides a very good approximation of the
atmospheric dynamics of a tidally locked terrestrial planet orbiting
an M dwarf. If we assume that the planet is (1) dry, i.e., that the
effects of moisture and clouds on the infrared emission to space are
small, and (2) the atmosphere is sufficiently thick, we find that a
WTG model could not easily be distinguished observationally from a
full 3D GCM calculation of the atmospheric dynamics using the MIRI
instrument on the JWST and assuming the planet orbits a nearby M dwarf
star. Although we did tune the WTG model to the GCM, the fact that
this is possible with only 3 WTG model parameters indicates that the
WTG approximation is an acceptable first approximation for
understanding dry Earth-like M dwarf planets.  It is reasonable to
expect that many planets in or near the habitable zone of M dwarfs
will be fairly dry due to reduced volatile delivery
\citep{Raymond:2007} and nightside ice trapping
\citep{2013arXiv1304.6472M}. If a measured thermal phase curve is
consistent with clouds having a minor or no effect, then a Monte Carlo
fit of the three-parameter WTG model to the observed thermal phase
curve should provide a good description of the dynamical behavior of
the atmosphere.

Of course the habitable zone planets that will be of most interest are
the ones that actually have liquid water at the surface on the dayside
in at least some regions, and will therefore tend to have clouds.  In
this case interpretation of the phase curve would require some
understanding of the cloud behavior \citep{2013Yang}. Since we have
shown that the WTG approximation is a good approximation of the
atmospheric dynamics, it should be possible to construct a
version of the WTG model incorporating cloud effects with only one or
two extra parameters that could  be fit to observed phase curves. In
such cases it would be beneficial to also run a GCM to confirm that
implied cloud behavior is reasonable, which demonstrates the value of
using a hierarchy of climate models to understand the climate of
exoplanets.

Aside from the direct implications of using WTG models to decipher the
basics of atmospheric dynamics and planetary climate from thermal
phase curves, it is important to emphasize that the fact that this
works reasonably well justifies the use of the WTG approximation in
theoretical studies, e.g., by \citet{2011ApJ...726L...8P} and
\citet{2011ApJ...743...41K}. This is extremely beneficial because of
the physical insight that the WTG approximation can provide into the
problem of understanding the climate of tidally locked
terrestrial planets.

\acknowledgements We thank R. Pierrehumbert for critical discussions
early in the development of this project, F. Ding for assistance with
the GCM, N.  Cowan and J. Bean for useful discussions of phase curve
observability, and K. Stevenson for insight into observational
applications of this work. We thank N. Cowan, D. Fabrycky, and an
anonymous reviewer for comments on a draft of this paper. This
research was begun with funding provided by a California Institute of
Technology Summer Undergraduate Research Fellowship. DSA acknowledges
support from an Alfred P. Sloan Research Fellowship.


\end{document}